# Assessment of slip transfer criteria for prismatic-to-prismatic slip in pure Ti from 3D grain boundary data

E. Nieto-Valeiras[a,b], E. Ganju[c], N. Chawla[c], J. LLorca[a,b,1]

[a]*IMDEA Materials Institute, 28906 Getafe, Madrid, Spain*
[b]*Department of Materials Science, Polytechnic University of Madrid/Universidad Politécnica de Madrid, 28040 Madrid, Spain*
[c]*School of Materials Engineering, Purdue University, West Lafayette, IN 47907, USA*

**Abstract**

Slip transfer and blocking across grain boundaries was studied in a Ti foil with a strong rolling texture deformed in tension. Prior to deformation, the shape of the grains and the orientation of the grain boundaries were quantified through laboratory scale diffraction contrast tomography (LabDCT). Mechanical deformation led to the activation of <a> prismatic slip, and slip transfer/blocking was assessed in > 300 grain boundaries by means of slip trace analysis and electron backscatter diffraction. A categorical model was employed to accurately assess slip transfer, and the "F1 score" of various slip transfer criteria proposed in the literature was evaluated for the first time from 3D grain boundary information. Remarkably, for the prismatic-dominated slip transfer in the current Ti sample, the results show that the best predictions of slip transfer/blocking are provided by the angle $\kappa$, which is directly related to the residual Burgers vector, and by the Luster-Morris parameter $m'$. In contrast, metrics based on the twist angle $\theta$ and on the LRB criterion were not able to predict accurately slip transfer/blocking. Thus, the extensive analysis of the 3D grain boundary data and the novel application of LabDCT was able to help clarify the role of grain boundary orientation on the mechanisms of plastic deformation in polycrystals with strong prismatic-dominated slip.

[1] Corresponding author
*Email address:* javier.llorca@upm.es, javier.llorca@imdea.org (J. LLorca)

## 1. Introduction

Grain boundaries (GB) play a critical role in the deformation of polycrystals. The lack of continuity between atomic lattices at the GBs hinders dislocation slip and leads to the formation of dislocation pile-ups and stress concentrations. The nature of the GB controls the degree of slip transmission and is also responsible for size effects associated with grain size [1, 2], crack nucleation as a result of GB embrittlement because of hydrogen [3, 4], irradiation [5, 6], or cyclic loading [7, 8], as well as the nucleation of deformation twins [9, 10]. Slip transfer across the GB may also occur depending on the degree of misorientation and several geometric criteria have been proposed over the years to assess whether slip transfer or blocking will occur at a given GB [11].

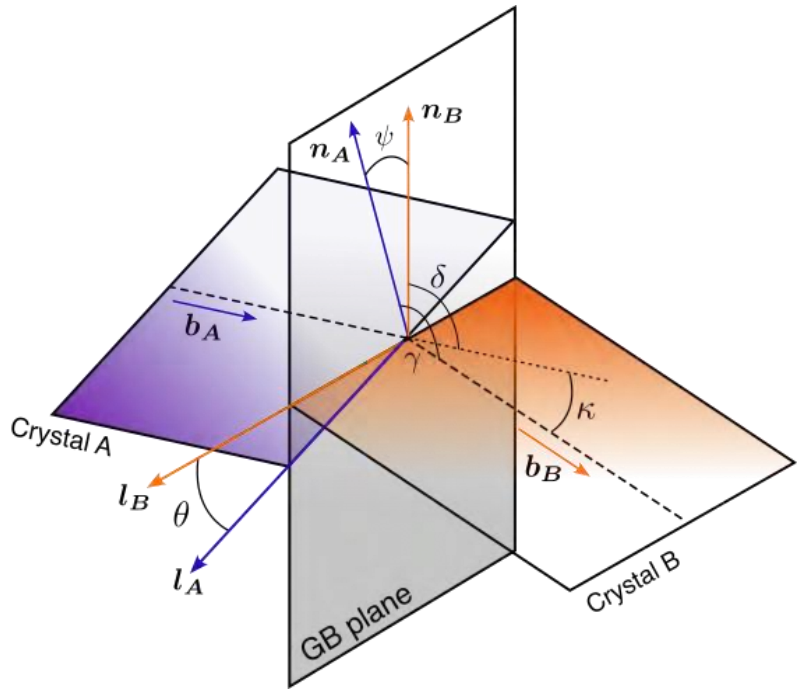

**Figure 1:** Schematic representation of the geometrical alignment between incoming and outgoing slip systems across a GB. Adapted from the work of Bieler *et al.* [11]; the version used in [13] has been modified in this work to include the angles $\delta$ and $\gamma$, and the vectors $l_A$ and $l_B$.

The occurrence of slip transfer across a GB has been studied extensively and it has been associated with good geometrical alignment between the incoming and outgoing active slip systems [12]. Figure 1 shows a detailed description of the slip transfer problem. The incoming slip system is characterized by the Burgers vector $b_A$ and the slip plane normal $n_A$, whereas the outgoing slip system is determined by the Burgers vector $b_B$ and slip plane normal $n_B$. Finally, $l_A$ and $l_B$ stand for the traces of the incoming and the outgoing slip planes within the GB plane.

The main angles that describe the alignment/misalignment between incoming and outgoing slip systems across the GB are $\kappa$, the angle between Burgers vectors, $\psi$, which is the angle between slip plane normals, and $\delta$ and $\gamma$, which stand for the angles between the incoming Burgers vector and the outgoing slip plane normal and vice-versa, respectively. Finally, $\theta$ (often designated as the twist angle) represents the angle between the traces of the incoming ($l_A$) and outgoing ($l_B$) slip planes with the GB plane.

Based on these angles, various slip transfer criteria have been proposed in the literature. For instance, slip transfer is less likely to take place if the magnitude of the residual Burgers vector left at the GB, $\Delta b$, is very large [14, 15], which is associated with the angle $\kappa$

$$\Delta b = |b_A - b_B| = 2 sin(\kappa/2) \qquad (1)$$

Livingston and Chalmers [16] pioneered the use of geometrical criteria to predict the likelihood of slip transmission across GB in pure Al bicrystals. They proposed a parameter $N$:

$$N = \cos\psi \, \cos\kappa + \cos\gamma \, \cos\delta. \qquad (2)$$

Slip transfer was likely to occur from a given incoming slip system to a certain outgoing slip system for $N$ of the largest outgoing slip system, with bounds between 0 and 1. They stated that the macroscopic continuity of the material at the GB requires the operation of additional slip systems in the vicinity of the boundary. Thus, under a pure shear stress approximation, the outgoing slip systems are activated by the stress concentrations due to dislocation pile-ups in the incoming slip system. The criterion was successfully used to predict the active outgoing slip system across a GB with a given incoming or primary slip system.

A few decades later, Lee et al. [17] proposed a combined criterion as an extension of what Shen *et al.* [18] published a year before. They stated that slip transfer will take place under three conditions: (1) the twist angle $\theta$ should be a minimum, (2) the resolved shear stress acting on the outgoing slip system should be maximized, and (3) the residual Burgers vector of the dislocation left at the GB ($\Delta b$) and, thus, the angle $\kappa$, should be minimized. Mathematically, the *LRB* criterion was expressed as:

$$LRB = \cos\theta \, \cos\kappa \qquad (3)$$

Later, Luster and Morris [19] proposed another criterion which assumes that slip transfer will occur when the compatibility factor $m'$, defined as:

$$m' = \cos\psi \cos\kappa \qquad (4)$$

is close to 1.

Experimental validation of the different slip transfer criteria has been conducted, in recent years, by electron backscatter diffraction (EBSD). This is typically done in combination with slip transfer/blocking evidence at GBs from slip trace analysis on the surface of deformed polycrystals imaged in a scanning electron microscope (SEM) [20-25]. In general, slip transfer is associated with low misorientation angles between adjacent grains at the GB, and also with low values of the residual Burgers vector, $\Delta b$, and high values (close to 1) of the $m'$ compatibility factor. In these studies, however there are many cases where a non-trivial number of slip blocking events where the theories predict slip transfer.

The two dimensional (2D) experimental investigations presented above have several limitations that hinder their capability to assess accurately the effect of GB orientation on slip transfer. Firstly, the active slip system cannot be uniquely identified when multiple slip systems share the same slip plane, as the {111} slip system family in FCC materials or the {0001} basal slip systems in HCP materials. Traditionally, the active slip system was identified as that with the highest Schmid factor, but this is not always the case, as has been recently demonstrated in Mg [26]. Secondly, EBSD is a surface technique that cannot provide information about the geometry of the GB across the thickness of the samples. Thus, the effect of the twist angle $\theta$ and the effectiveness of LRB (which directly depends on the GB normals) as a slip transfer criterion cannot be accurately assessed.

Some experimental investigations have revealed that the $\theta$ angle may affect the occurrence of slip transfer across GB even though the slip systems (between incoming and outgoing grains) are well aligned. For instance, Lee *et al.* [27] performed a TEM-based study in which it was observed that the outgoing slip plane was that with the smallest twist angle in order to accommodate slip in FCC materials. In the case of Σ3 annealing twin boundaries in Ni, the twin plane normal can be directly calculated. As reported by Gennée *et al.* [28] and Nieto-Valeiras and LLorca [25], $\theta \approx 0°$ appears as a necessary condition for slip transmission across coherent twin boundaries. In HCP materials, slip blockage was observed in a few GBs in pure Mg where $m'$ was close to 1 but $\theta > 60°$, as recently reported by Sarebanzadeh *et al.* [26]. They measured the GB orientation within the sample by sequential milling with a focused ion beam. This technique has also been used in other investigations [29, 30] but it is destructive and very time consuming, limiting the number of GB that can be analyzed.

In the last decade, laboratory-scale diffraction contrast tomography (LabDCT) has become a powerful tool to perform non-destructive high-fidelity 3D grain mapping in metallic polycrystals [31]. Notably, the accuracy of LabDCT has been extensively validated, against electron backscatter diffraction (EBSD) data [32, 33, 34], and hence it can help assess the role that the 3D geometrical characteristics of the grains and grain boundaries play in the slip transfer mechanism.

In this investigation, for the first time, we were able to obtain all the necessary geometrical parameters, through a combined application of LabDCT, EBSD, and SEM, and accurately measure and quantify slip transfer behavior using 3D grain orientation and grain boundary data. We employed EBSD and LabDCT methodologies to generate grain maps of Ti foils, which were subsequently exposed to tensile deformation. The LabDCT data was especially beneficial, enabling us to visualize the three-dimensional geometry of grain boundaries. By combining the 3D grain maps from LabDCT with SEM based slip trace analysis, we were able to determine the actual active prismatic slip systems and examine slip transfer phenomena at several hundred GBs in our Ti sample. This sizable and unique dataset, encompassing both GB orientation in 3D and 2D slip transfer characteristics, was used to assess the efficacy of different slip transfer criteria proposed in the literature. This work is the first to provide a comprehensive understanding of the impact of geometrical characteristics of the grains and grain boundaries on slip transfer and plastic deformation in metals using a synergistic application of nondestructive 2D and 3D characterization techniques.

## 2. Materials and experimental procedure

High purity (99.99%), cold-rolled Ti foils (0.25 mm in thickness) were used in this study (Goodfellow, Huntingdon, United Kingdom). A microtensile dog-bone sample with a uniform central region of 5×1 $mm^2$ was machined from the foil by electro-discharge machining. The sample was heat-treated at 850°C for 8 hours in a tubular vacuum furnace under argon atmosphere to minimize oxidation. The sample was gently ground with 1200 grit paper to remove the fine oxidation layer on the surface after the heat treatment. Afterwards, it was electropolished to mirror finish on both surfaces of the central gauge section using a Struers A3 electrolyte at 27 V and room temperature.

Electron backscatter diffraction (EBSD) maps were acquired from both surfaces of the gauge length of the sample. The crystallographic orientations were obtained in a FEI Helios NanoLab 600i dual-beam microscope with an Oxford Instruments EBSD detector. The whole gauge length was mapped with several overlapping maps followed by stitching in AZtec. The EBSD maps were acquired at 20 kV and 2.7 nA, with a step size of 3 $\mu$m, in order to accurately resolve the microstructural features. The overall quality of the EBSD maps was above 95% indexing. The maps were post-processed with the MTEX Matlab Toolbox [35] with a grain boundary misorientation threshold of 2° to capture low-angle grain boundaries.

The 3D grain structure of the materials was mapped using lab-scale diffraction contrast tomography (LabDCT) (Zeiss Versa 620 X-ray microscope, Dublin, CA). The sample was positioned at equal distance from the X-ray source and the detector (14 mm), keeping the Laue focusing geometry, and a source voltage and power of 160 kV and 15 W, respectively, were used. A square aperture with a 750×750 μm opening was used to illuminate the sample and a 4X DCT detector with a transmission beam stop was used to capture the diffraction spots during the scan. A conventional DCT scanning approach was carried out and the crystal orientations in the sample were illuminated in the angular range of 0° to 360° while keeping the sample at a fixed height. This consists of 181 diffraction projections acquired in angular increments of 2°, for each vertical position. In order to map the full gauge length in the sample with the selected aperture, seven vertically overlapping scans were "stitched" (Figure 2 (b)), resulting in a total scan time of 63 h. The acquired DCT projections were iteratively reconstructed with GrainMapper3D (Xnovo Technology) using a forward modeling approach [32] until the average completeness values reached at least 90%, where completeness is defined as the ratio of observed diffraction spots to forward simulated diffraction spots. The reconstructed grain map from LabDCT had a resolution of 5 $\mu$m/voxel (voxel is a 3D pixel). Post-processing of the grains was carried out using Dream3D [36] to segment the volume into grains and grain boundaries. ParaView [37] was used for visualization purposes. More information about the DCT acquisition strategy and the post-processing workflow, as well as the accuracy of the technique to map a similar Ti sample can be found in [34].

The sample was tested in uniaxial tension in a Kammrath and Weiss microtensile testing machine equipped with a 1 kN load cell at a quasi-static strain rate of $\approx 10^{-3} s^{-1}$ to a strain of about 2%. The combination of SEM imaging and grain orientation before deformation by EBSD was used to identify the active slip systems in the grains after deformation. The active slip system identification by conventional slip trace analysis is based on the comparison between the orientation of the experimentally observed slip bands with that obtained from the crystal orientation, as measured by EBSD [22, 23, 25]. It is worth mentioning that the crystal orientations provided by DCT could have also been used instead of EBSD to determine the theoretical orientation of the slip traces. However, the spatial resolution of EBSD at the sample surface is better than that of DCT, improving the accuracy in the case of very small grains.

## 3. Results

### 3.1. 3D grain and grain boundary characterization

The DCT data reconstructed in GrainMapper3D was imported into Dream3D [36] to characterize the microstructure in 3D. Each voxel contained information about the Euler angles and the completeness value attained after the forward-modeling process. The data were denoised using a minimum completeness value of 0.1 and a misorientation threshold of 2° for grain segmentation. Any spurious data were filled using an erode/dilate process with two iterations. Figure 2(a) shows the reconstructed sample gauge after post-processing in Dream3D, where the left half shows the voxelized volume. The sample was composed of approximately 24 million voxels and 588 grains, and the information about grains' positions, sizes, shapes, and misorientations was exported for the 3D microstructural analysis. A detailed view of two neighbor grains reconstructed with 5 $\mu m$ voxel size is provided in Figure 2(c).

Once the volume of the sample was reconstructed, a triangular surface mesh was generated over the inner (GBs) and outer surfaces of the sample. The initial triangular surface mesh was smoothed using a Laplacian algorithm to minimize any stair-stepped appearance, and the resulting triangle data (triangle id, centroid, area, curvature, misorientation, and normal direction) were exported for further analysis. The right half of Figure 2(a) shows the generated surface mesh, composed of roughly 1.5 million triangles and 2319 GBs, where the outer surfaces of the sample have been removed for the sake of GB visualization. Figure 2(d) shows the surface mesh at the GBs of the grains in Figure 2(c), where the shared GB normals have been plotted as yellow arrows.

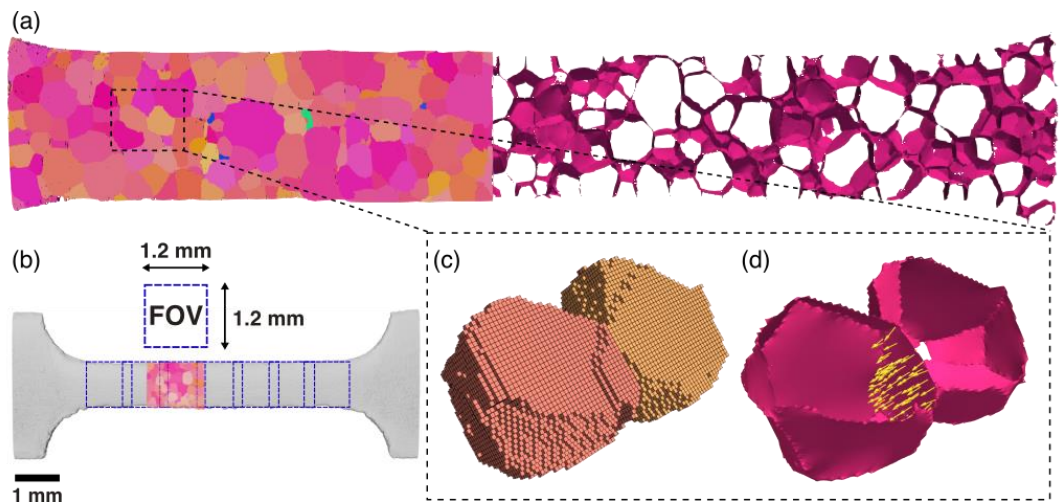

**Figure 2:** Microstructural characterization by DCT. (a) Reconstructed DCT map of the sample gauge colored according to the inverse pole figure with respect to the screen

plane (color key in Figure 3 (b)). The left half corresponds to the reconstructed volume of the sample, and the right half corresponds to the surface mesh generated at the GB. (b) Sample scanning strategy consisting of 7 overlapping sub-regions to map the full gauge length. (c) Detailed view of two neighboring grains (dashed squared area in (a)), where the reconstructed voxelized volume is shown in (c), and the corresponding surface mesh is shown in (d). In the latter, some GB surface normals are drawn in yellow.

The normal vector to each triangle in the surface mesh was directly obtained from the surface mesh, and the GB normal direction, $n_{GB}$, was estimated. To this end, a different subset was generated for every GB from the triangular mesh, so that only the corresponding triangle normals were selected. The triangle winding was carefully checked and corrected so that all normal directions were pointing in the same direction. Given that the mean curvature of the studied GBs is close to 0, the GB normal was taken as the mean of the triangle normals included in the surface. The calculated GB normals, together with the observed active slip systems at both neighboring grains, allowed for the calculation of the twist angle $\theta$ according to Figure 1

$$\theta = tan^{-1}\left(\frac{||l_A \times l_B||}{l_A \cdot l_B}\right) \quad (5)$$

where $l_A$ and $l_B$ stand for the traces of the incoming and the outgoing slip planes with the GB plane, respectively, (Figure 1) that can be calculated as follows:

$$l_A = n_{GB} \times n_A$$

$$l_B = n_{GB} \times n_B \quad (6)$$

The lognormal grain size distribution obtained from DCT data is plotted in Figure 3(a). The average grain size is 111 ± 58 $\mu$m. The grain boundary misorientation angle (also provided by DCT data after grain segmentation) is plotted in Figure 3(b) and presents a bimodal distribution with peaks at $\approx 15°$ and at $\approx 70°$.

The GB inclination $\beta$ is defined as the angle between the GB plane, characterized by the GB normal $n_{GB}$ (obtained from LabDCT), and the plane perpendicular to the sample surface and oriented along the length of the sample. Thus, vertical GB (perpendicular to the sample surface) are characterized by high $\beta$ (close to 90°). The GB inclination distribution $\beta$ is plotted in Figure 3(c). As observed in Figure 2(a), there is a significant fraction ($\approx 20\%$) of vertical or close to vertical GBs with $\beta > 80°$, that grew very likely during the heat treatment due to the small thickness of the sample. Nevertheless, approximately 55% of the GB have an inclination < 65°.

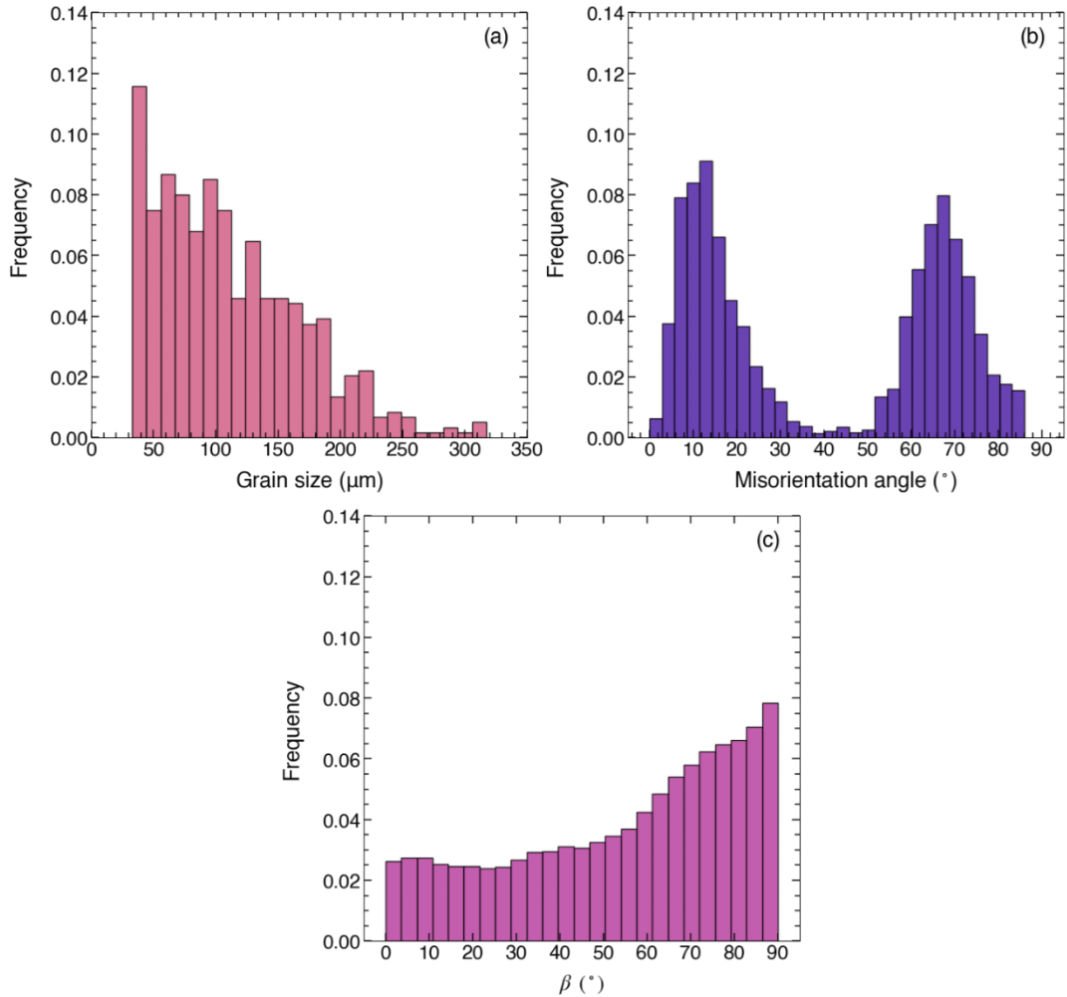

**Figure 3:** Microstructural features of the Ti sample obtained by DCT. (a) Grain size distribution, (b) Grain boundary misorientation distribution, (c) Grain boundary inclination ($\beta$) distribution.

### 3.2.   2D microstructural characterization

The microstructure of the surface of the sample obtained by EBSD is shown in Figure 4. The grains have been colored according to the inverse pole figure perpendicular to the surface, and the color key is provided by inverse pole figure color code in Figure 4(b). The raw EBSD map displayed in Figure 4(a) shows a good indexing rate with equiaxed grains throughout the gauge of the sample. The post-processed EBSD map of the rectangular region on the left side of the sample is shown at higher magnification in Figure 4(c). The GB are colored according to their misorientation angle, ranging from very low misorientation GB (blue) to high misorientation GB (yellow). The relative orientation of the hexagonal unit cells is shown in the center of each grain and reveals a marked texture. The basal plane pole figure in Figure 3(d) shows a strong inclined basal texture, with the *c*-axes tilted about ±30° from the Z axis (perpendicular to the surface). This texture is typical of cold-rolled commercially pure Ti [38], and agrees to the bimodal misorientation angle distribution obtained from the DCT data (Figure 3(b)). As a result of this texture, tensile deformation will be favored by the activation of two of the three prismatic <a> slip systems in the hexagonal unit cell, whose Schmid factors are close to 0.5 (Figure 4(c)).

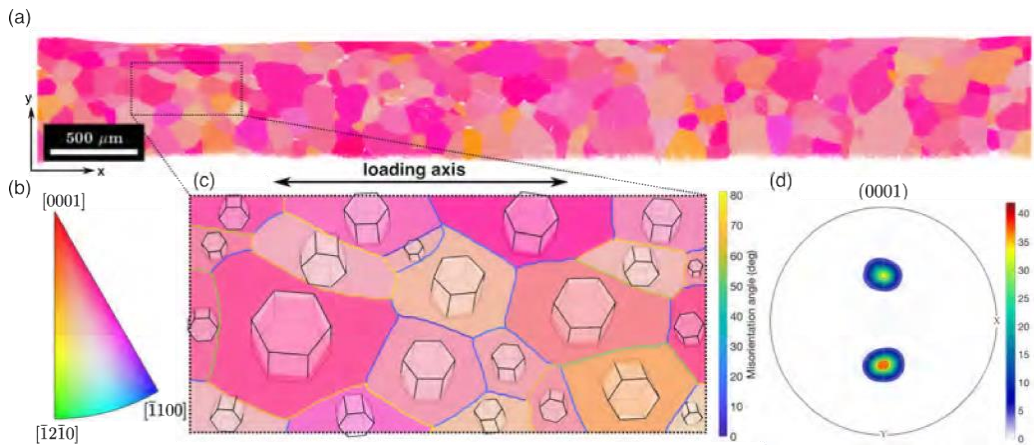

**Figure 4:** Microstructure of the gauge section of the sample from EBSD. (a) Raw EBSD map colored according to the inverse pole figure perpendicular to the surface in (b). White dots correspond to regions that were not indexed by EBSD. Loading axis is horizontal, as indicated by the arrow. (c) Post-processed EBSD map of the rectangular region indicated in (a) at higher magnification. GB are colored according to their misorientation angle in degrees, and the orientation of the hexagonal unit cell is shown at the center of each grain. (d) Basal plane (0001) pole figure. Intensity is marked in multiple of random distributions.

### 3.3. *Slip trace and slip transfer analysis*

The analysis of 233 grains in the SEM revealed that only prismatic <a> slip took place in the sample, favored by the strong texture. The lack of activation of other basal <a> or pyramidal <c+a> slip systems facilitates the unique identification of the active slip system (slip plane and direction) since the three prismatic planes lead to slip traces with different orientation on the sample surface. It should be noticed that the identification of the active slip system through conventional slip trace analysis is not effective if the slip plane or the Burgers vector is parallel to the sample surface. To avoid this problem, the material and texture were carefully selected for this investigation to limit the number of active slip systems to avoid the noise and the complexity introduced by the activation of many (and different) slip systems [45]. In particular, (1) the prismatic planes are always oblique to the loading direction, and only one or two of them were suitably oriented for slip, and (2) the basal planes (which may be parallel to the sample surface) are very poorly oriented for plastic deformation given the geometry of the hexagonal lattice. Thus, it is very unlikely that other slip systems (besides prismatic) could be active, and any slip traces different than prismatic were not observed throughout the sample. The low CRSS of prismatic slip in pure Ti and the only observation of prismatic slip traces after deformation seem to indicate that prismatic slip is enough to accommodate the plastic deformation in the sample.

Moreover, the deviation between the theoretical slip trace (calculated from the grain orientations obtained from the EBSD of the undeformed sample) and the observed slip trace was always <5°. These small deviations between the observed and theoretical slip traces can be explained by the change in crystal orientation caused by plastic deformation of the sample. Once one active prismatic slip system was impinging onto a GB, a number of parameters were measured for that GB. In the case of multiple slip in one or both grains across a GB, the best combinations in terms of the geometrical alignment between the active slip systems were chosen. This means that, in case that double prismatic slip is observed at both sides of a GB like in the case of Fig. 5 (c), among the four possible combinations between the slip systems, the two that provide the lowest $\kappa$ and $\psi$ are chosen. The recorded parameters include GB misorientation, the Schmid factors of the incoming and outgoing slip systems across the GB, and the four angles that characterize the geometrical alignment between the incoming and outgoing slip systems at the GB on the surface of the sample ($\kappa$, $\psi$, $\delta$, $\gamma$) (Figure 1). This information was recorded for 361 GB considered reliable due to their similitude between EBSD and DCT, and was used to assess the slip transfer criteria.

The active prismatic slip systems were identified on either side of each GB, and they were classified according to the occurrence or absence of slip transfer in three distinct groups. The first group included those GBs in which slip transfer was convincingly observed, as revealed by the visible and unequivocal continuity between all or most of the slip traces at both sides of the boundary (Figure 5(a) and (b)). In some cases, only one slip system was active in both grains across the GB (Figure 5(a)) while double slip transfer along two different pairs of slip systems took place in other cases (Figure 5(b)). Slip transfer in these cases was confirmed by the absence of any changes in surface topography at the GB, indicating that deformation between the neighboring grains was homogeneous.

The second group included those GB where slip blocking was observed, or in other words, slip transfer was not observed (Figure 5(c)). The slip traces were stopped at the GB or did not match at both sides of the GB and, sometimes, feather-shaped micro-volumes of deformation appeared near the boundary, indicating the presence of stress concentrations associated with heterogeneous plastic deformation [14, 15, 28]. The presence of a ledge or significant change in the topography at a GB is also indicative of heterogeneous deformation at both sides of the GB, and it is often associated with uncertain or poor slip transfer conditions [11, 22].

The third and last group corresponds to what we define as partial slip transfer (Figure 5(d)). These cases show slip transfer to some extent but were different from the perfect slip transfer cases depicted in Figure 5(a) and (b). For instance, just a few slip bands matched across a GB or some traces matched but then faded away a few micrometers away from the GB.

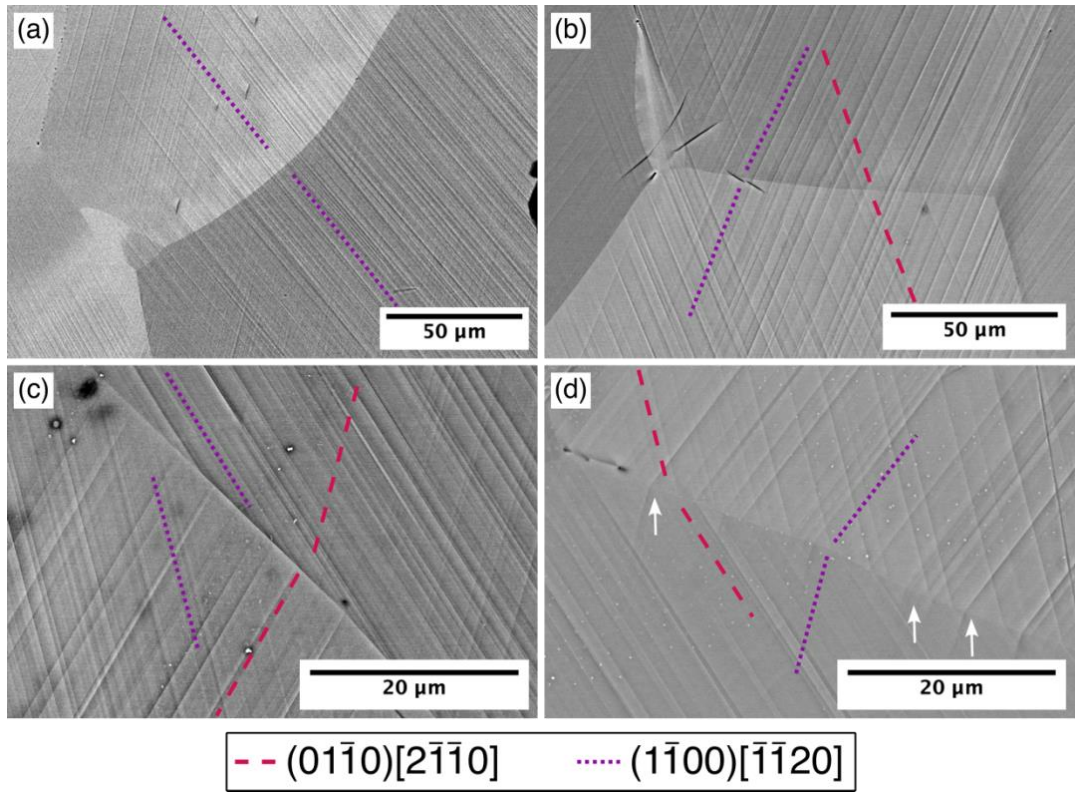

**Figure 5:** Prismatic-to-prismatic slip transfer/blocking across GB. The theoretical orientation of the slip traces of the active slip systems are indicated by the dashed lines. (a) Slip transfer. (b) Double slip transfer. (c) Slip blocking and ledge at the GB. (d) Partial slip transfer. The arrows indicate slip bands fading towards the grain interior.

Out of the 362 analyzed GB, convincing slip transfer was observed in 43.9% and partial slip transfer in 11.6% of the cases, leaving a remaining 44.5% of slip blocking instances (see Figure 6(a)). Therefore, the likelihood of slip transfer in this sample is distributed in an approximately 50-50 fashion, i.e., 50% for partial + full slip transfer and 50% for slip blockage. The relationship between slip transfer, partial slip transfer, and slip blocking with GB misorientation angle is plotted in Figure 6(b). It shows that slip transfer (either perfect or partial) occurs at low misorientation GBs (< 30°). Nevertheless, there are still slip blocking events located at low misorientation GB, suggesting that the GB misorientation angle is not sufficient to predict prismatic-to-prismatic slip transfer across GBs in pure Ti. With a comprehensive database of the geometrical characteristics of the grains and grain boundaries, we can now assess how the different slip transfer criteria perform for the current Ti sample.

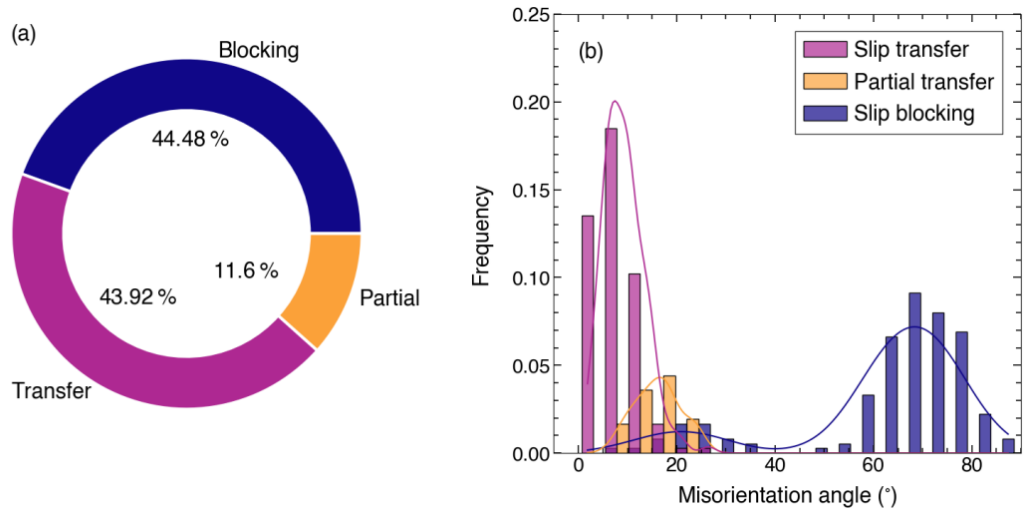

**Figure 6:** Distribution of slip transfer events in the microstructure: (a) Classification of prismatic-to-prismatic perfect slip transfer (purple), partial slip transfer (orange) and slip blocking (dark blue) across GB in Ti, and (b) Influence of GB misorientation angle distribution on the occurrence of perfect slip transfer, partial slip transfer and slip blocking.

*3.4. Evaluation of slip transfer criteria*

The slip transfer criteria presented in the introduction rely on the angles $\kappa$, $\psi$, $\gamma$, $\delta$ and $\theta$, in addition to the misorientation angle. All of them can be determined for each GB from the EBSD information of the neighboring grains except for $\theta$, which requires 3D information about the orientation of the GB within the sample, given by $n_{GB}$. This information was determined from the DCT data for 266 GB out of the 362 analyzed by slip trace analysis (74%). This difference is due to the limitations of the DCT technique to capture accurately the shape and orientation of small grains [34]. Only GBs whose shape was accurately determined by both LabDCT and EBSD, were included in the analysis. The twist angle $\theta$ was calculated using Eqs. (5) and (6), which take as input the grain boundary normal, $n_{GB}$, obtained from LabDCT data, and the normals of the active slip planes across the GB, $n_A$ and $n_B$, obtained from the combination of SEM imaging and EBSD crystallographic information.

Traditionally, $\kappa$ (that determines the residual Burgers vector) and $\psi$ (that captures the misorientation between the incoming and outgoing slip planes) have been considered the most relevant angles to assess slip transfer [11, 21], following the Luster-Morris parameter $m'$. Their ability to predict slip transfer is shown in Figure 7(a), which shows that the majority of slip transfer events were found when $\kappa$ and $\psi$ were < 20°, in agreement with previous investigations [22, 25, 23, 26]. In these cases, the incoming and outgoing slip systems are well aligned because the magnitude of the residual dislocation left at the GB is small, and the resolved shear stress acting on the incoming slip system is easily transmitted to the outgoing slip system because $\psi$ is also small. Moreover, all partial slip transfer instances are located at low to moderate values of these angles (< 30°), and all the blocking events are shifted towards higher angles, except for three outliers where there is a ledge at the GB, indicating that plastic deformation at these GB was not homogeneous. Two out of these three outliers correspond to the GB shown in Figure 7(c), which presents a significant change in topography between the adjacent grains. In this case, two slip blocking instances are observed between both pairs of active slip systems (purple-to-purple and pink-to-pink) despite that they are suitably aligned for slip transfer. The information provided by the conventional slip trace analysis technique is not sufficient to explain why slip transfer is not observed across these GBs. Nevertheless, other techniques, such as high-resolution digital image correlation (HRDIC) could be employed to assess the heterogeneous distribution of the deformation and the effect of the local stress on the occurrence of slip transfer.

The occurrence of slip transfer/blocking is plotted in Figure 7(b) as a function of $\kappa$ and $\theta$. Perfect slip transfer is associated with low values (< 20°) of both angles but no instances of slip transfer are found if $\kappa$ > 30°, regardless of $\theta$. In fact, low values of $\theta$ only lead to slip transfer if they are associated with low values of $\kappa$. Moreover, many instances of perfect or partial slip transfer are found if $\kappa$ < 25° even for very large values of $\theta$. Finally, the occurrence of slip transfer/blocking is plotted in Figure 7(c) as a function of $\psi$ and $\theta$. Slip transfer is mainly associated with low values of $\psi$ but there is no obvious trend with $\theta$. Moreover, it should be noted that large values of $\theta$ (> 30°) can be found even when the incoming and outgoing slip planes are well aligned and $\psi$ < 20°. Theoretically, when $\psi$ is small the slip planes are well aligned, and thereby $\theta$ is very small and independent of the GB inclination angle $\beta$ (Figure 1). However, as $\psi$ increases, the twist angle $\theta$ is more sensitive to the GB inclination. Thus, it cannot be concluded that –in general– good alignment between the active slip planes leads to low twist angles and it is necessary to obtain this information experimentally. Overall, these results seem to indicate that $\theta$ plays a secondary role to determine slip transfer/blocking, as compared with $\kappa$ and $\psi$ for the current prismatic-dominated slip transfer. This can be an effect of the strong texture in the sample, where all plastic deformation can be accommodated through prismatic slip. The influence of $\theta$ in a randomly oriented sample is still not well understood and could affect slip transfer between prismatic and non-prismatic slip systems.

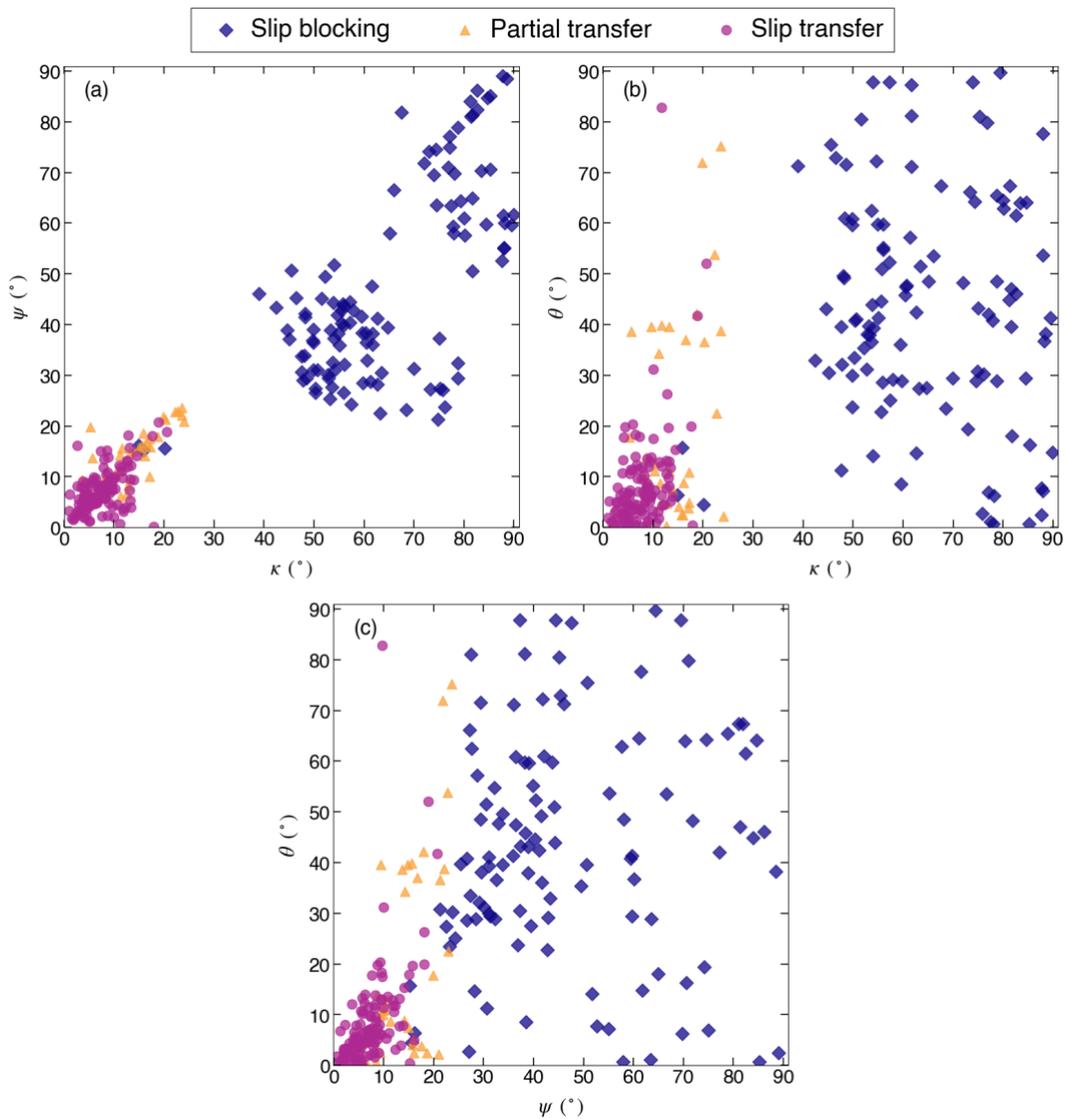

**Figure 7:** Dependence of slip transfer on $\kappa$, $\psi$, and $\theta$: (a) Influence of $\kappa$ and $\psi$ on the occurrence of slip transfer/blocking across the GB. (b) Influence of $\kappa$ and $\theta$ on the occurrence of slip transfer/blocking across the GB. (c) Influence of $\psi$ and $\theta$ on the occurrence of slip transfer/blocking across the GB.

In order to obtain more quantitative information, a categorical model was used to assess the influence of the different angles ($\kappa$, $\psi$, $\theta$, $\delta$ and $\gamma$ in Figure 1, together with the GB misorientation calculated from the crystal orientation of the neighboring grains) on the likelihood of slip transfer. To this end, all the data were divided in two categories, namely slip transfer (including partial slip transfer) and slip blocking. This is a reasonable assumption, since partial slip transfer instances do transfer slip to some extent. The occurrence or absence of slip transfer for each GB was characterized by the categorical variable *slip* that can only take the values of 1 or *true* (slip transfer) or 0 or *false* (slip blocking).

The next step was to determine the optimum value of each angle that is able to discriminate slip transfer from slip blocking for each angle. The procedure is schematically shown in Figure 8(a) for $\kappa$. For any value of $\kappa$ (such as $\kappa$ = 45° in Figure 8 (a)), the data set will be divided in two groups, with *slip* = 1 (slip transfer) to the left of $\kappa$ and *slip* = 0 (slip blocking) to the right of $\kappa$. This leads to a binary categorical model, where the data can be arranged in a confusion matrix including the true positives (TP), true negatives (TN), false positives (FP) and false negatives (FN), as depicted in Figure 8(b). Different model classification metrics, such as *Precision* (Eq. 7) and *Recall* (Eq. 8) [39, 40], can be defined as

$$Precision = \frac{TP}{TP + FP} \quad (7)$$

$$Recall = \frac{TP}{TP + FN}. \quad (8)$$

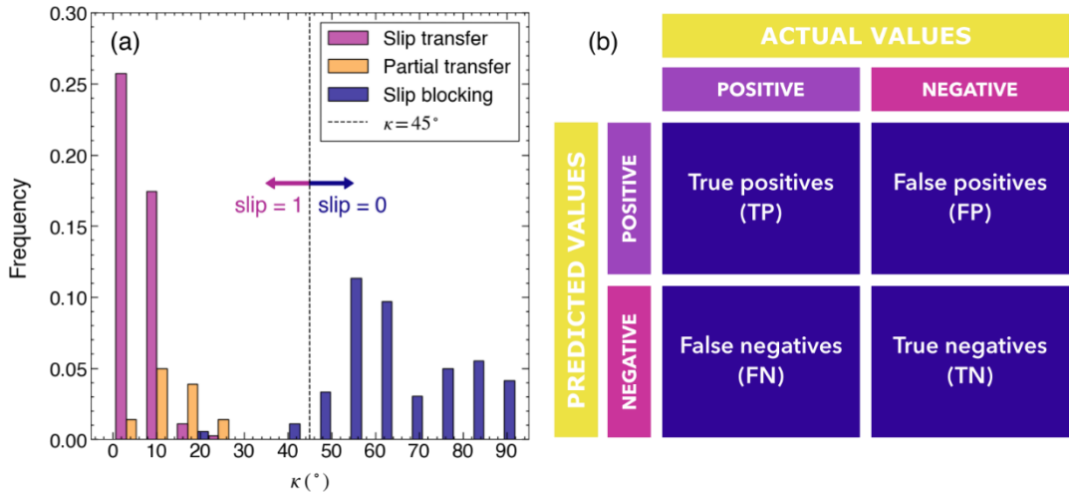

**Figure 8:** (a) Influence of angle κ on the occurrence of slip transfer, partial slip transfer and slip blocking. A black vertical dashed line is drawn in κ = 45° to represent a potential threshold that divides the data in predicted slip events (*slip* = 1) to the left of the line and predicted blocking events (*slip* = 0) to the right of the line. (b) Schematic representation of a confusion matrix to compare the actual and predicted values slip transfer/blocking for any categorical variable.

*Precision* or confidence deals with the "Predicted positive" row of the confusion matrix, determining how accurate the model is in predicting the positive outcomes out of all the outcomes predicted to be positive. *Recall* or sensitivity, deals with the "Actual positive" column of the confusion matrix, and measures the completeness of the positive predictions or how accurately the model was able to identify the positive outcomes out of all positive outcomes that were actually present. In other words, *Precision* indicates the accuracy of the positive predictions, whereas *Recall* refers to the percentage of positive values correctly classified by the model. In a model with 100% *Precision*, there are no false positives, and every positive prediction is correct, whereas with 100% *Recall* there are no false negatives, and every negative prediction is correct.

Any of these metrics can be used to evaluate the performance of any categorical model, depending on the specific purpose of the model. However, they have to be combined into a single metric when both precision and recall are given the same importance. This is normally achieved through the F1 score [40]

$$\text{F1} = 2\frac{Precision \cdot Recall}{Precision + Recall} \qquad (9)$$

that can be interpreted as a measure of overall model performance ranging from 0 (worst) to 1 (best).

The threshold value for slip transfer for all the angles that characterize the GB was assessed from the maximum value of F1 and the results are summarized in Table 1. They show very similarly low values for the GB misorientation, $\kappa$, $\psi$ and $\theta$, and very similarly high values for $\delta$ and $\gamma$. However, the predictions given for each angle are not similarly accurate, the higher the optimum F1, the better this angle is able to predict slip transfer/blocking. This information is summarized in the F1 score matrix (Figure 9), that shows the maximum value of F1 for each angle in the diagonal of the matrix as well as the maximum F1 that can be obtained by choosing any pairwise combination of two angles with their respective thresholds presented in Table 1. The F1 score has been used as a heat map variable and therefore the colors of each box in the matrix represent the magnitude of F1 as compared to the other boxes.

According to the data in Figure 9, the best indicator of slip transfer is $\kappa < 24.5°$, with F1 = 0.993. The F1 values for all the out-of-diagonal terms are smaller than this value, which means that the combination of two angles does not lead to a better predictive performance for the current microstructure. It is also worth mentioning that the combination of three or more angular parameters leads to worse results than those presented in Figure 9. The next best performances are attained by the pairwise combinations of misorientation, $\kappa$ and $\psi$, whereas the performance of criteria based on $\theta$ angle always attain the poorest performances to predict slip transfer/blocking. It should be noted that these findings may differ for a more random microstructure and further exploration is needed to assess more diverse cases.

Table 1: Threshold angle (according to the F1 score) to predict slip transfer.

| *Angle* | *Threshold angle* |
|---|---|
| GB misorientation | < 24.5° |
| $\kappa$ | < 24.5° |
| $\psi$ | < 23° |
| $\theta$ | < 23° |
| $\delta$ | > 74° |
| $\gamma$ | < 78.5° |

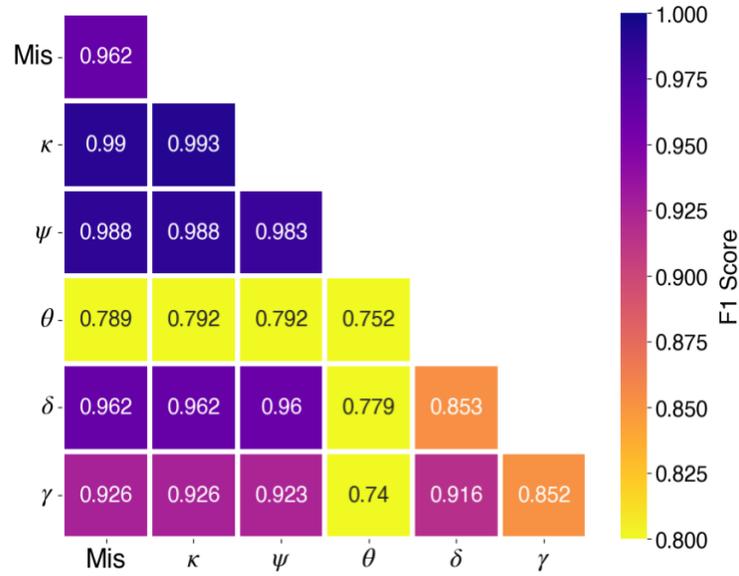

**Figure 9:** F1 score matrix for slip transfer/blocking. The diagonal terms represent the performance of a single angle in terms of F1 score whereas the out-of-diagonal terms represent the F1 score obtained by the combination of two angles. The F1 score value is indicated in each box and is also represented as a heat map variable, ranging from yellow (lowest) to blue (highest). The threshold that leads to the maximum F1 for each angle is depicted in Table 1.

A similar analysis was carried out to assess the performance of the slip transfer criteria presented in the introduction of this paper, namely GB misorientation angle, Luster-Morris parameter $m'$, normalized Burgers vector $\Delta b/b$, *LRB*, and *N*. The corresponding threshold values of each parameter for slip transfer that led to the highest F1 score are shown in Table 2. These thresholds were used as the best possible separation between the slip transfer and blocking populations, and the corresponding F1 score matrix is presented in Figure 10.

Besides low misorientation angles, slip transfer is likely to occur for low $\Delta b/b$ and high $m'$ and *N*. The accuracy of this thresholds to predict the slip transfer/blocking, according to the F1 score, is shown in Figure 10. The best performance is obtained with either $\Delta b/b$ or $m'$, that lead to F1 = 0.993. Other slip transfer criteria (*N* and misorientation angle) lead to lower F1 scores while the *LRB* criterion gives the lowest F1 metric, indicating that it is not a good criterion to predict prismatic-to-prismatic slip transfer/blocking at GB in Ti. The combination of two slip transfer criteria (for instance, $\Delta b/b$ and $m'$, $\Delta b/b$ and *N*, $m'$ and *N*) improves the F1 score up to the values obtained with only $\Delta b/b$ or $m'$ but not more. On the contrary, the F1 score of *LRB* or of any combination of *LRB* with another criterion is much lower, indicating that *LRB* is not a reliable criterion for slip transfer/blocking for this particular system.

Table 2: Optimum threshold (for maximum F1 score) for different well-known slip transfer criteria.

| Parameter | Threshold |
|---|---|
| GB misorientation | < 24.5° |
| $m'$ | > 0.8 |
| $\Delta b/b$ | < 0.45 |
| LRB | > 0.65 |
| N | > 0.9 |

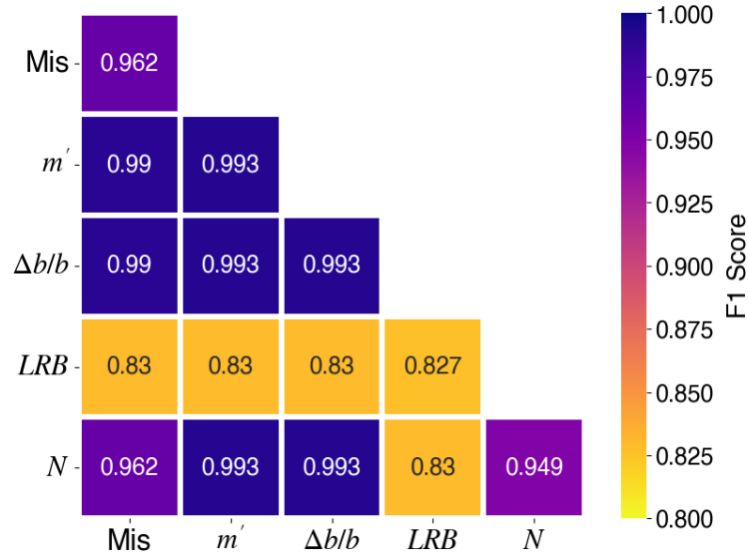

**Figure 11:** F1 score matrix for slip transfer/blocking. The diagonal terms represent the performance of a single slip transfer criterion in terms of F1 score whereas the out-of-diagonal terms represent the F1 score obtained by the combination of two slip transfer criteria. The F1 score value is indicated in each box and is also represented as a heat map variable, ranging from yellow (lowest) to blue (highest). The threshold that leads to the maximum F1 for each slip transfer criterion is depicted in Table 2.

## 4. Discussion

The results presented in the previous section provide - to the authors' knowledge - the first assessment of slip transfer across GB in polycrystalline metals in which a) the active slip systems across the GB are actually identified; b) the GB geometry is fully characterized in 3D and c) slip transfer/blocking data are obtained in a statistically significant number of GBs. These unique advantages have allowed us to robustly identify the governing geometrical parameters that impact the slip transfer in our prismatic slip dominated Ti sample.

One of the primary conclusions of the analysis of the data is that the normalized residual Burgers vector across the GB, $\Delta b/b$ appears to be the best criterion to predict slip transfer, which is likely to occur if $\Delta b/b < 0.45$ (or $\kappa < 24.5°$) while slip blocking will be dominant otherwise. Similar scores are obtained in the case of $m'$, which includes information about the residual Burgers vector and about the angle $\psi$ between the normals of the incoming and the outgoing slip planes, which are related to the resolved shear stresses in both planes.

It should be noted, however, that the particular texture of the Ti sample leads to the presence of two prismatic planes in each grain with very high Schmid factors. Thus, if $\Delta b/b$ is small and there is not a large barrier for slip transfer, the incoming dislocations are likely to find another slip system with a high resolved shear stress in the neighbor grain that will be available to accommodate slip transfer. Thus, it is possible that $m'$ (that includes the information about the residual Burgers vector) becomes a better predictor of slip transfer/blocking in other microstructures. In general, these results agree with previous investigations based on surface observations in different polycrystalline metals [11, 21, 22, 25, 23].

The second main conclusion of the investigation is that the twist angle $\theta$ (and, thus, the *LRB* criterion) does not appear to be a good predictor of slip transfer/blocking for the grain orientations in prismatic slip dominated Ti sample tested here. Many instances of perfect and partial slip transfer were found for high values of $\theta$, and slip transfer never occurred for low values of $\theta$ unless the residual Burgers vector was also small (Figure 7(b). This result is to some extent surprising because the limited information available in the literature seemed to indicate the $\theta$ played a major role in slip transfer. Zhou *et al.* [41] studied basal-to-prismatic and basal-to-pyramidal slip transfer in a Mg alloy. The actual basal slip system was determined using lattice-rotation analyses and slip transfer was reported in 23 cases, which corresponded to high $m'$ in 16 cases and low $\Delta b/b$ in 21 cases. The GB orientation was measured by sectional grinding in one of the cases with slip transfer and low $m'$ and it was found that $\theta = 26°$. Nevertheless, $\theta$ was not measured (but estimated) in other cases with slip transfer and low $m'$ and solid conclusions could not be reached. Besides that, low $\Delta b/b$ seems to be a very good predictor of basal-to-prismatic and basal-to-pyramidal slip transfer in Mg.

Furthermore, Sarebazadeh *et al.* [26] recently carried out a detailed analysis of basal-to-basal slip transfer in pure Mg. The actual active basal slip system in each grain was carefully identified by means of slip trace - modified lattice rotation analysis [42] and slip blocking when $m' < 0.45$ and $\Delta b/b < 0.6$ was only found in a few cases, that were analyzed in detail. They corresponded to situations in which basal slip was almost parallel to the GB (and, thus, slip lines did not cross the boundary) or to five cases in which the GB orientation was measured through-the-thickness through sequential focused ion beam. $\theta > 60°$ in all these cases and it was concluded that high twist angles may hinder basal to-basal slip transfer in Mg. In fact, the actual slip transfer mechanism for high twist angles is not known and it has been suggested [43] that it is necessary the activation of secondary slip systems (such as <a> basal or pyramidal slip in Ti) near the GB to accommodate slip transfer between the primary prismatic slip systems when $\theta$ is high. While this is possible in Ti, because the critical resolved shear stress for <a> basal slip is not far away from that for <a> prismatic slip, this is unlikely to happen in Mg, in which the critical resolved stress for <a> basal slip is much lower than for other slip systems, blocking slip transfer for high twist angles. Zhou *et al.* [41] measured subsurface dislocation activity by differential aperture x-ray microdiffraction in five grains and found that <c+a> dislocations were active, but their activity did not appear at the surface. This observation implies that other unexpected dislocation sources may be assisting the slip transfer process under the surface.

In the case of special GB such as twin boundaries, the effect of the twist angle can be assessed because the GB normal can be directly determined from the known crystallographic rotation between the parent and the twin. Genée *et al.* [28] analyzed slip transfer across Σ3 annealing twin boundaries in a Ni-based superalloy and reported that the $\theta$ was the main geometrical parameter determining slip transmission across twin boundaries, where $\theta = 0°$ appeared as a necessary condition for slip transfer. Chen *et al.* [44] performed uniaxial tensile testing at 650°C in Inconel 718 Ni-based superalloy, and also reported that $\theta = 0°$ and $\Delta b < 0.6b$ are sufficient conditions to ensure slip transfer across twin boundaries. However, the actual pair of active slip systems across the twin boundary was dictated by the maximum Schmid factor and were not conclusively established. Nieto-Valeiras and LLorca [25] acquired a significant number of slip transfer/blocking occurrences across grain and twin boundaries in a pure Ni sample deformed in tension. Slip transfer across twin boundaries was only observed when $\Delta b \approx 0$, regardless of the twist angle $\theta$, and pointed out that slip transfer across annealing twins seems to be controlled by dislocation cross slip. Regarding regular grain boundaries, the effect of $\theta$ was not assessed due to the lack of information about the GB normal, but slip transfer was generally observed when $m' > 0.8$.

It should be finally noted that - given the experimental difficulties to measure the GB orientation through the thickness of polycrystalline samples, some studies have assumed that GB are perpendicular to the sample surface [28, 44] to study the effect of $\theta$ on the likelihood of slip transmission. Under this assumption, Genée *et al.* [28] determined that slip transfer was mostly observed for $\theta < 30°$, although it was also found for higher twist angles $> 40°$. Similarly, Chen *et al.* [44] stated that slip transfer was likely to occur when LRB > 0.82, but several slip blocking instances were found at high *LRB* values while slip transfer was also observed below such threshold. However, this hypothesis may lead to large errors in the evaluation of $\theta$, as indicated by [28].

Thus, the experimental information obtained in this investigation provides -for the first time- a sound evidence that prismatic-to-prismatic slip transfer across regular GB in Ti is favored by a minimum magnitude of the residual Burgers vector and also by the good alignment between the incoming and outgoing slip planes, while the twist angle plays a secondary role in the process. It is important to mention, however, that the negligible effect of the twist angle in this case could be affected by the strong texture present in the sample, which could lead to easier plastic deformation accommodation. The influence of the twist angle could still be important for slip transfer between non-prismatic slip systems in randomly oriented Ti samples.

This investigation showed that some of the simple geometrical slip transfer criteria (i.e., those based in Δb and m') can be used to assess slip transfer/blocking in > 99% of the GBs and, thus, are suitable to be incorporated into crystal plasticity models to simulate the behavior of polycrystals. In addition, that large twist angles do not hinder prismatic-to-prismatic slip in Ti. However, there are a few GBs within the analyzed dataset whose behavior do not follow these geometrical criteria. These outliers could be caused by the local stress distribution in the GB neighborhoods, which may change the local Schmid factors, hindering slip transfer. Conventional slip trace analysis does not provide any information about the strain heterogeneity within the sample and other techniques such as high-resolution digital image correlation would be required in order to assess the effect of the local stress state around GBs on the likelihood of slip transfer. Further investigations, using the methodology outlined in this paper, will help clarify whether these conclusions can be extended to other slip systems in other metallic alloys.

## 5. Conclusions

In this paper, we analyzed slip transfer across grain boundaries by means of SEM based slip trace analysis and electron backscatter diffraction in a thin Ti sample which presented a strong rolling texture. The two-dimensional microstructural analysis was complemented with three-dimensional grain maps, procured through LabDCT, which allowed for the determination of 3D grain shapes and the orientation of grain boundaries within the sample. After a thorough characterization, the sample was loaded in tension which resulted in the activation of its <a> prismatic slip systems. The orientation of the active prismatic slip system(s) in each grain were determined from the orientation of the slip traces, and slip transfer/blocking was analyzed in > 300 grain boundaries on the sample surface. We assessed the likelihood of slip transfer/blocking as a function of the main angles that characterize the incoming and outgoing slip systems and the grain boundary (misorientation angle, $\kappa$, $\psi$, $\theta$, $\gamma$ and $\delta$) by means of the "F1 score" within the framework of categorical models. The analysis was further extended to assess the efficacy of widely used slip transfer criteria available in literature (misorientation angle, $\Delta b/b$, $m'$, $LRB$ and $N$). Via a thorough analysis of the various geometric characteristics of grains and grain boundaries, we arrived at the following observations:

(1) For the current prismatic slip dominated Ti sample, the best angle to predict slip transfer was the angle between the incoming and outgoing slip directions, with $\kappa < 24.5°$ as the threshold for slip transfer with an F1 score of 0.993.

(2) The next best performances were attained by the pairwise combinations of grain boundary misorientation angle, and the angles between incoming and outgoing Burgers vectors and slip planes, $\kappa$ and $\psi$.

(3) Interestingly, the predictions for slip transfer based on the twist angle $\theta$ resulted in the poorest performance for the current prismatic slip dominated system.

(4) From the slip transfer criteria available in literature, the best performance was obtained with either $\Delta b/b$ or $m'$ slip transfer criteria (that lead to an F1 score of 0.993) while other slip transfer criteria (based on $N$ and or the misorientation angle) led to lower F1 scores, with the $LRB$ criterion giving the lowest F1 metric.

(5) The combination of two slip transfer criteria (for instance, $\Delta b/b$ and $m'$, $\Delta b/b$ and $N$, $m'$ and $N$) improved the F1 score up to the values obtained with only $\Delta b/b$ or $m'$, but not more.

(6) In contrast, any combination of $LRB$ with another criterion led to much lower F1 scores, indicating that $LRB$ was not a reliable criterion to predict prismatic-to-prismatic slip transfer/blocking at GBs in the current Ti sample.

The results of this research highlight the effectiveness of combining established characterization techniques like EBSD and SEM with newer methods like LabDCT, in thoroughly decoding the mechanisms of slip transfer in metals. The study has delivered substantial evidence that, in the case of the prismatic-slip dominated Ti sample, prismatic-to-prismatic slip transfer across regular GB in Ti is facilitated by a minimal residual Burgers vector and a favorable alignment between the incoming and outgoing slip planes. The methodology demonstrated in this research could assist in shedding light on the comparative influence of the different geometrical parameters in more diverse slip systems.

**Acknowledgements**

This investigation was supported by the project (MAD2D-CM)-IMDEA Materials funded by Comunidad de Madrid and by the Recovery, Transformation and Resilience Plan and by NextGenerationEU from the European Union, and by the María de Maeztu seal of excellence from the Spanish Research Agency (CEX2018-000800-M).

**Data availability**

The experimental data generated in this investigation can be accessed at [Link to be added upon acceptance of the paper].